# GROMATICS' ILLUSTRATIONS
# FROM NEWLY DISCOVERED PAVEMENTS IN POMPEII

**Massimo Osanna [1], Giulio Magli[2], Luisa Ferro[2]**

[1]*Università degli Studi Federico II-Parco archeologico di Pompei*
[2]*Politecnico di Milano*

**ABSTRACT**

In recent archaeological investigations of Pompeii' Regio V houses,  a series of new, rather enigmatic images on the floors of the entrance and of the Tablinium of the so-called House of Orion have been unearthed.
These images are presented here together with a tentative interpretation, which is based on a clear analogy of their content with the illustrations of the codexes of the Gromatics, the Roman agrimensores.

1. INTRODUCTION

As part of the *Great Pompeii Project*, inaugurated in 2014 and co-financed by the European Community, a series of safety measures were taken in various areas of the ancient town. In particular, capillary interventions throughout the city were made, with special attention to the hydro-geological risks associated with the unexcavated embankments present in the archaeological area. These interventions included excavations in Regio V, precisely in a portion of the city enclosed between *Vicolo delle Nozze d'Argento* and *Vicolo di Cecilio Giocondo*, for a total of more than 1000 square meters.

Since these new archaeological investigations have been announced (on March 27, 2018, exactly 270 years after the "discovery" of Pompeii) many discoveries have followed, adding valuable information to the history of the city. The work was carried out by completing the excavation of *Vicolo delle Nozze d'Argento* and by clearing *Vicolo dei Balconi*, called in this way due to the discovery of five balconies (Osanna 2017, 2019).

Two houses open on the western side of the newly excavated street. Taking the road that has just emerged from the lapilli, after a few tens of meters on the left you come across a house with a solemn, ancient facade. To this house, which has been brought to light in his completeness only in December 2018, the name Casa di Orione has been given. This is because of the two extraordinary mosaics found inside, probably representing Orion's myth and catasterism in the rooms and rather enigmatic images on the floors of the entrance and of the Atrium have been unearthed.

The aim of the present paper is to give a description and a tentative interpretation of such images (Osanna, 2018; Osanna being published)

2. THE HOUSE OF ORION AND NEW EXCAVATIONS

A small section of the House of Orion was already excavated between 1892 and 1893 (Mau 1894). At that time, August Mau discovered a portion of a garden with a three-column peristyle and parts of three other small columns, probably belonging to a loggia on the upper floor. The excavation was also documented by photographic images (F.lli Esposito, 1890 c.a., Osanna 2016). He also found, in the northern enclosure of the garden, a frame in the form of an aedicule containing a painting depicting Jupiter.

At the time of the discovery, it was thought that these finds actually belonged to a house that stood on the Via di Nola (still called Casa di Giove). Today's intervention has instead shown that the complex belongs to a house with entrance along *Vicolo dei Balconi*. It is a central *Atrio* house surrounded by rooms that appear as official (as opposed to private) halls. These rooms have a rich decoration in the first style, with stucco panels imitating marble slabs (*Crustae*) in red, black, yellow, and green color and with stucco frames with serrated moldings. The atrium was probably completed by a Doric stucco frieze (fragments of which remain) with a blue and red finish.

The houses *in Atrium,* corresponding to *Cava Aedium* used by Vitruvius' *De Architectura* and Varrone's *De Lingua Latina,* are based on a architectural scheme is very close to that of the Samnite House. The house opens up from the road in a small *Vestibulum*. The project of the interior is geometrically quite rigorous: the atrium has a quadrangular shape with central axis aligned with the entrance door and the vestibule. In the background, a small garden is surrounded by boundary walls, with a peristyle developed along one side only. Along the longer sides of the Atrium the rooms (Cubicola) are placed with a *trinarian* partition, that is, three modular spaces arranged on each side. Very recently, in two of these rooms two mosaics have been discovered which with all probabilities represent Orion's hunting Myth (north-est side) and Orion's Catasterism (south-west side). It is the first time that a Catasterism is found explicitly represented in the Roman iconography, and this is a strong hint into an interest in (and perhaps knowledge of) astronomy of the owners of the house, a thing we shall come back on later.

The *Tablinium* is off-center with respect to the main axis; its plan was likely reduced to make room for an open space allowing air and light to filter from the garden to the Atrium. The Atrium in itself is the center of the whole house composition. The *Cava aedium* was built geometrically, as is the case for other Domuses in Pompeii like the House of the Faun, Pompeii VI, 12, 1-8 (Hoffmann, 1996; Dwyer, 2001; Ferro 2018). Starting indeed from the line of the outer wall facing the Fauces up to the line of the inner wall of the Atrium towards the Peristilium, the rectangular plan of the Atrium corresponds to two golden rectangles whose larger sides are equal to the shorter side of the Cava aedium itself. The accuracy by which these are obtained is impressive and there is seldom any doubt that they were intended in the project. Indeed we remind that golden rectangles are obtained if the proportions $A-A^1 : A^1-B = AB : A-A^1$ are close enough to the irrational number $\phi$ (which of course, being irrational, cannot be exactly obtained). Our laser scanner measures of the house furnish a value for $\phi$ correct to four digits (1,6179 to be compared with 1.6180).

The building which can be seen today is the result of two main constructive phases. In particular, the images which are the main object of the present study are located on a floor that constitutes a layer superimposed on a previous floor, belonging to a Tuscan Cava aedium with *Impluvium* which was obliterated in the new building phase.

The removal of the Impluvium shows that substantial changes were made when the house was restored. Actually, the building changed from a house in *Cava aedium tuscanicum* (that is, the oldest and most frequent house in Pompeii, with a central opening called Compluvium) to a "basilica" House with *Cava aedium testudinatum.* The ceiling was completely closed by a vaulted roof with only a front opening, and the center of the atrium was highlighted with a stone, obliterating the water drain system which was not needed any more.

### 3. THE IMAGES ON THE NEWLY DISCOVERED PAVEMENTS

The floor belonging to the second-phase is in the so-called *Cocciopesto*. This was a cement mixture with crushed terracotta fragments (sometimes substituted by crushed black lava fragments, and in this case called Lavapesta); as rustic as it may seem today, it was actually a very durable material composed of several layers and made by specialized workers. (Maiuri, 1943).

Already in Republican times, *Cocciopesto* pavements were embellished with decorative motifs. These decorations were obtained by placing small stones or mosaic tiles of different colors. Many examples of this kind of artworks are known from the Roman world, including Pompeii itself (Meleagro's house). In the case of the Domus object of this study the *Cocciopesto* is decorated with figures composed of fragments of mosaic arranged in the last layer – like "seeds" - and then rolled so as to incorporate the tesserae into the compound without altering the figures (a similar technique is visible in the Fauces of the Domus of the Faun (Pompeii VI, 12, 1-8), in which the inscription *Have* is also printed, as a welcome greeting for those entering the house).

At least as far as the present authors are aware, however, all the Roman cocciopesto decorative motifs recovered so far are laid out according to simple, repeated geometric patterns, such as lozenges, concentric squares, or the recently found svastica-like decoration of the Roman villa of Montegibbio, Modena. The decorations under exam are instead non-repeated, specific works of art placed in an otherwise uniform pavement, and as such they are the unique known so far. There are 3 such images. The one we shall label (IM 1, FIG. 3), is located in the atrium, almost on the entrance of the central court; the other two, labelled (IM 2, FIG. 4) and (IM 3, FIG. 4) respectively, are located on the opposite side of the court, almost on the same axis of the first.

Looking at the images, the complete originality with respect to all other known *Cocciopesto* decorations is immediately apparent. First of all, as mentioned, there is no repeated pattern or any other motif in the rest of the *Cocciopesto* pavement's surfaces of the house, which were left undecorated and uniform in color. This singles out the 3 images and shows that they were conceived and realized with a precise intention, in suitably chosen points of the house. Before describing their content, therefore, it is worth analyzing their placement in the context of the house plan. The first image (IM 1) is found in the terminal part of the Entrance Fauces shortly before entering the atrium. The second and the third (IM 2 e IM 3) are located in the passage room that connects the atrium to the garden. Their disposition was carried out taking into account the geometry of the atrium with a somewhat precise calculation. The golden ratio of the Atrium was indeed the starting point of the arrangement: the first image (IM 1) is placed starting from the long side of the golden rectangle of the Atrium at the intersection with the axis of the house. The project was with all probabilities developed in terms of Oscan feet (1 O.F=0.275 cm) for instance the distance from the intersection of the lines constituting the center of the first image circle up to the center of the second image circle (IM 2) measures 11.53m, to be compared with 42 O.F=11.55 m. The IM 2 and IM 3 symbols are not arranged along the axis of the atrium identified by the IM 1 image but displaced to the south-west of it. In this way their line of sight aligns with the perspective of a person entering from the street (Vicolo dei Balconi).

The center of the IM 2 circle and the point D (belonging to the golden rectangle ABCD) constitute the vertices of a right-angled triangle (in the figure colored in orange) with base coinciding with the transverse axis of the atrium.

Let us now turn to what is represented in the images.

IM 1 Represents a square inscribed in a circle, with two sides of the square extending for a few tiles out of the circle. The circle (diameter 37cm) is cut by two perpendicular lines, one of which coincides with the longitudinal axis of the atrium.

From the arrangement of the mosaic tiles, one can easily recognize the inscribed square; the corners of it point (roughly) to the cardinal points. The side of the square also coincides with the long side of the golden rectangle. The first element/image therefore appears as a sort of rose of the winds that relates the orientation of the house (which is 44° to the north of east, from inside looking out) to the cardinal points, and identifies a regular division of the circle in eight equally spaced sectors.

Interestingly, the drawing finds close correspondence in a attempt to a reconstruction of a (lost) image of Vitruvius' text made by Rusconi in the 16 century (*Dell'Architettura secondo Vitruvio*, 1590).

IM 2 It is a relatively complex image. A circle (diameter 35cm) with an orthogonal cross inscribed in it is connected by five dots disposed as a sort of small circle to a straight line, at the end of which a further orthogonal segment appears. The circle is generated by two perpendiculars diameters with an apparently random orientation but which appears to be set at 30° relating to the transverse axis of the atrium.

IM 3 Is the simplest of the three images: five almost parallel and almost equally spaced segments which start orthogonally from a common baseline.

4. **INTERPRETATION OF THE IMAGES**

In what follows we are going to propose that these 3 images all refer to land surveying, a duty carried out by the Roman *mensores*. For this reason, we need to recall a few basic information about them and about their writings.

*4.1. The Roman surveyors and the groma*

As is well known, a fundamental role in the foundation of Roman colonies and in the division of the farmlands since very early times was played by land surveyor's –mensores or agrimensores – whose main working instrument was the so-called Groma. For this reason, the authors of technical treatises on land surveying are called "gromatic writers" or *gromatici*. We shall briefly recall here a few fundamental facts about them (Dilke, 1962).

The groma is known from reliefs on tombstones (of L. Aebutius Faustus at Ivrea, and of Popidius at Nocera) and from the discovery of the metallic parts of one of such instruments occurred in

Pompeii (Della Corte, 1912). A groma is composed in this way. First of all, a cross made of four perpendicular arms each bringing a hole at the end. Through these holes hang four cords with identical weights, acting as plumb-lines. The cross is not directly mounted over a pole, as instead one could infer from the 2-dimensional depictions on the tombstones. In fact, a fundamental part of the instrument, which is never represented in the reliefs probably for the difficulties in rendering the perspective, is a horizontal arm. At one end of the arm a pivot allows the center of the cross to rotate freely, while the other end is in itself free to rotate on the summit of the main pole. The length of the pole, which was made of perishable material and it is not conserved, is not known but it must have been not much different from the height of the tripod of a modern theodolite (say 3 or 4 Roman feet). The pole was fixed on the ground trough a bronze lancet.

The presence of the horizontal arm, which allowed the cross to rotate around the pole whenever necessary, is the key to the almost incredible accuracy reached by the Roman *mensores*, because it allowed the surveyor to target distant reference points without having the pole in the field of vision. The surveyor could therefore align with extreme precision two opposite, very thin plumb-lines with reference poles held at various distances by assistants or fixed in the terrain, in the same manner as *palines* are used in modern surveying. And indeed, as occurs today with modern, optical theodolites, the accuracy increases if the number of the poles and the distance at which they are targeted increases. We actually have solid proofs that huge Roman land-survey projects, such as the tracing of the Via Appia and the centuration of the Pontine marshes occurred at the end of the 4th century BC, were carried out with errors of alignments less than 15' (Magli et al, 2014, 2015).

### *4.2. The illustrations of the Corpus*

Fortunately enough, medieval codexes have passed on a collection of writings by the Gromatici, usually referred to as corpus agrimensorum. These codexes are not exempt of problems about dating, authorship and interpretation; generally speaking however, there is some consensus that the earliest ones should be the treatises by Sextus Frontinus, governor of Britain in 74 AD. Other early treatises should be those ascribed to a certain Hyginus (not to be identified with the known author of Augustan age), almost certainly written by more than hand. In any case what is especially important for us here is that the treatises are accompanied by illustrations (Dilke 1967). The oldest existing version of the corpus and the illustrations dates to the 6$^{th}$ century AD, but there are convincing proofs that the texts were copied by an original which was illustrated as well (Castagnoli, 1943). In other words, although the hand that actually draw the illustrations clearly is of the early Middle Age, their content is very probably close to the original one. So, the illustrations in the Corpus can be considered as "original" in this sense. They were apparently used for didactic aims, more or less like those of a modern technical manual. Most of the illustrations are connected with the centuriations of the fields and the placement of towns in the regularly divided landscape. The groma in itself is never shown; however, some images are related to the surveyor's techniques and in particular to orientation. In particular, there are two from Hyginus' treatise (FIG. 5) which are of special interest for us. The first shows the canonical division of space in eight regular sectors. This division is obtained by inscribing a square in a circle in such a way that the sides are parallel to the cardinal directions. The image relates to a context in which the author discusses different types of orientations (to the cardinal points or to the rising/setting positions of the sun). Orientation was of special relevance for the mensores, as the whole of their duties was shrouded with sacral aspects. In particular, they claimed for a direct descendance of their art from the sacred principles of *Etrusca Disciplina*. An adequate knowledge of astronomy was, therefore, clearly required to a mensor. The second image of interest shows schematically a regular division of land attached to a pre-existing division with a different orientation.

*4.3.   The possible meaning of the decorations*

The resemblance of Images 1 and 3 of the Orion House' pavement with the above mentioned two images of the Corpus discussed above is striking. Actually, the square+circle image is virtually identical. We are thus led to propose that this image, the first visible to a visitor of the house after the entrance, was a sort of icon of the connection of the owner with the principles, if not with the duties, of the mensores, and that image 3 represents schematically a land-surveying project. From this point of view a fascinating, although speculative, hypothesis arises. It can indeed be noticed that the average measure of the rectangles appearing in image 3 corresponds with a certain accuracy (Measures: Scaled drawing 0.86m x 0,11m; Actual Plot Regio I 259,8m x 34,4m) to the insulae of the urban addition to the city plan. This might of course be a pure chance, however the corresponding scale is worth noticing: it is 1:300. As far as we know there was no standard scale adopted by the Roman architects in their projects, however the 1:300 scale is the same scale of at least one among the most important Roman on-scale reproduction we know about, the Forma Urbis Romae. Further, the major line and the straight line on which parallel straight lines form a corner form a 97° angle. At this point it is necessary to make some further considerations. The orientation in ancient cartography and planning was not necessarily to the north, but referred to a monument or to the geographical element characterizing the place.The Forum and the parallel districts of via Mercurio in Pompeii, in particular, have a geographical orientation determined by the view of the Vesuvius Vulcan. But the main axis of Via Stabiana, via dell'Abbondanza and via Nola have different orientations, which is roughly 97°. Therefore if the axis of the house shown by the IMG 1 is rotated to coincide with Via Stabiana, the paths in IMG 3 superimpose with the lots of the last urbanization of Pompeii and the streets of Nola and Abbondanza.

All in all therefore the last image seems to correspond to a very schematic, planimetric representation of the most recent constructed part of the ancient city.

   Finally then, if we pursue the line of interpretation based on a possible "gromatics" content of the images, an explanation for image 2 almost naturally arises: it is the first known depiction of a *Groma*. The artist had the problem of representing a 3-dimensional object on a flat surface and with a relatively crude means of expression as *Cocciopesto*, but succeeded in showing the fundamental fact that the *groma* cross is detached from the pole apex. A difference with the unique known item of *groma* passed on to us is that the cross found in the excavations is not encircled. The possible presence of a circle in other items would not, however, influence the working of the instrument and is by far possible. In addition, representing a circled cross helped the artist in clarifying the structure of the object. Indeed, the cross-circle symbol was currently used by gromatici, and a *decussis* is indeed present on several cippi gromatici – stone cylinders used to signal the partition of the fields (see e.g. Tarasco). A symbolic intent could not be excluded, either. The cross-circle was related to the symbolism connected with the foundation of cities and colonies since Etruscan times, as shown, for instance, by the cross-inscribed pebbles found buried at the crossing of main streets in the Marzabotto excavations and, more recently, discussed in other Etruscan contexts (Bagnasco, 2019).

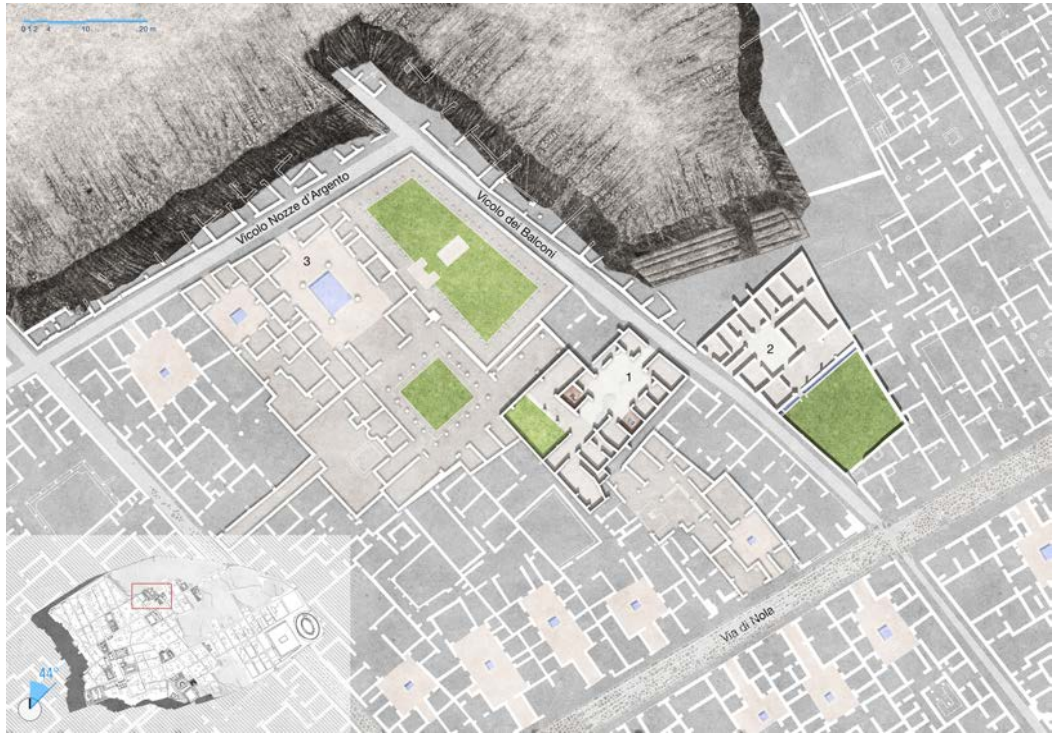

**Figure 1.** Plan of Regio V (1. The Domus of Orion, 2. The Domus of the Garden)

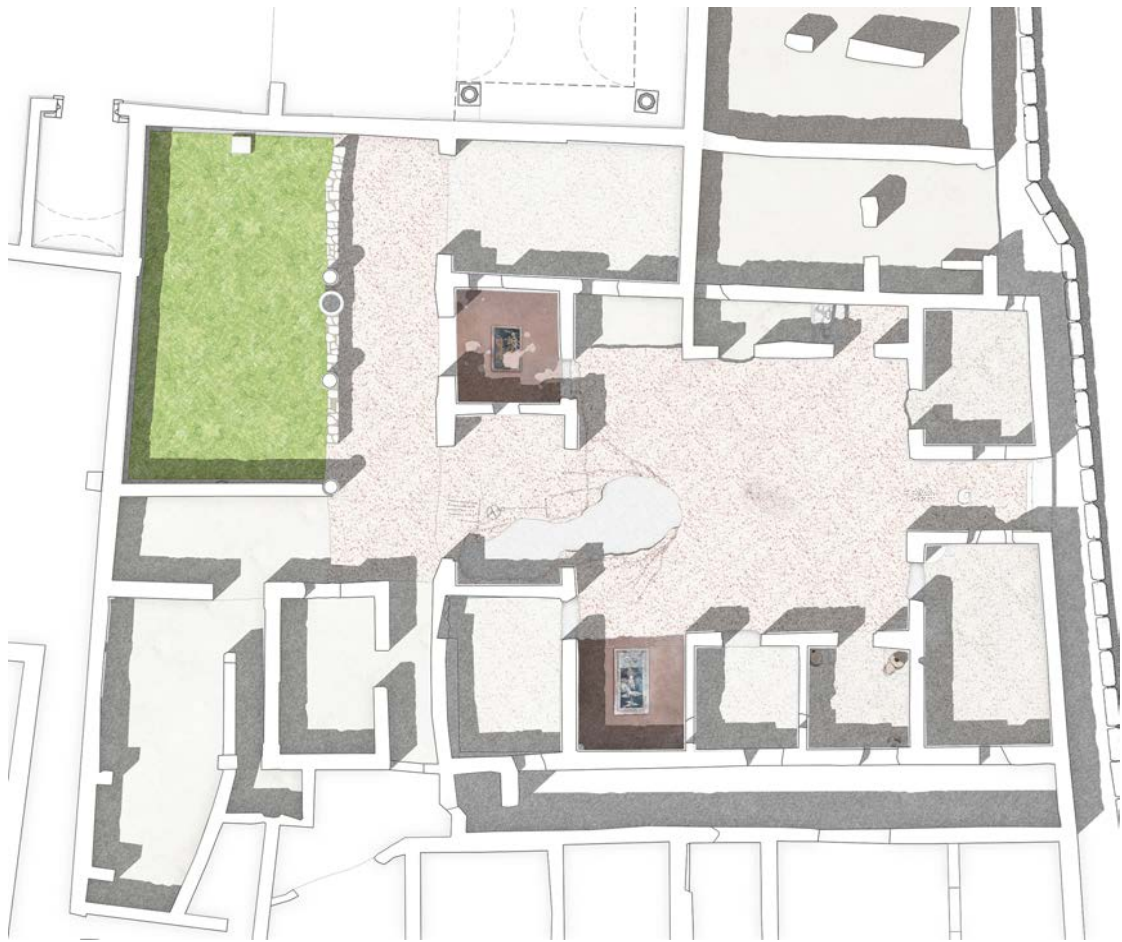

**Figure 2.** The Domus of Orion. Plan

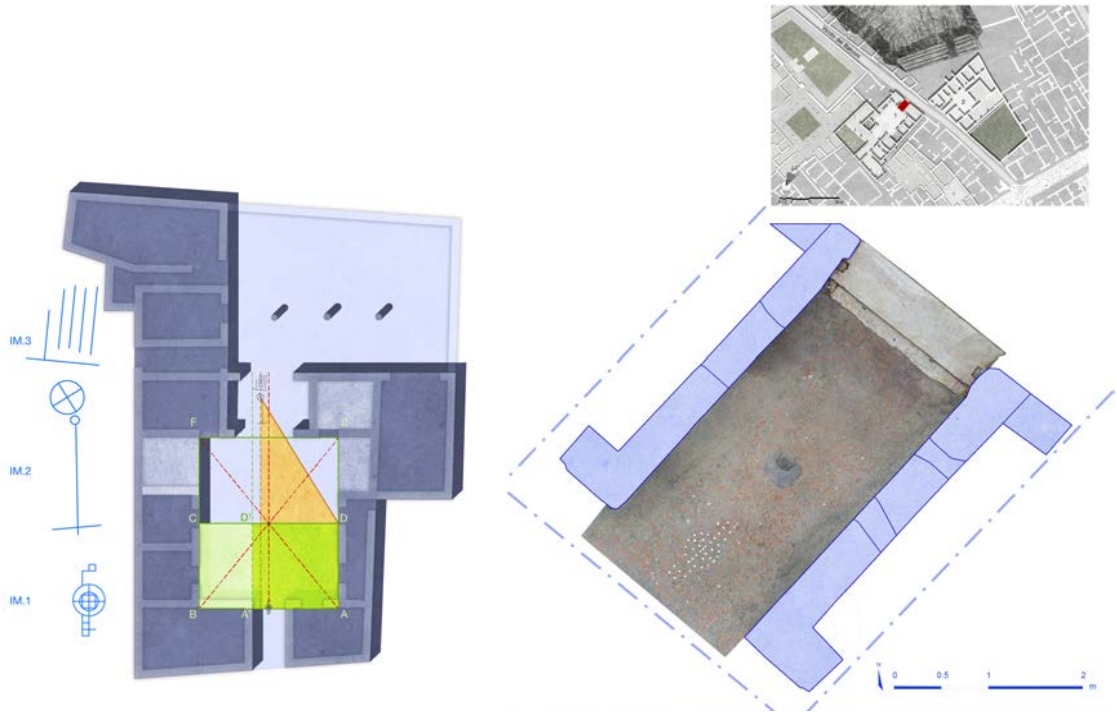

**Figure 3.** Left. Plan of the *Domus* with the geometric study of the *Atrium*. Right The *Vestibulum* with IM1, photo plan.

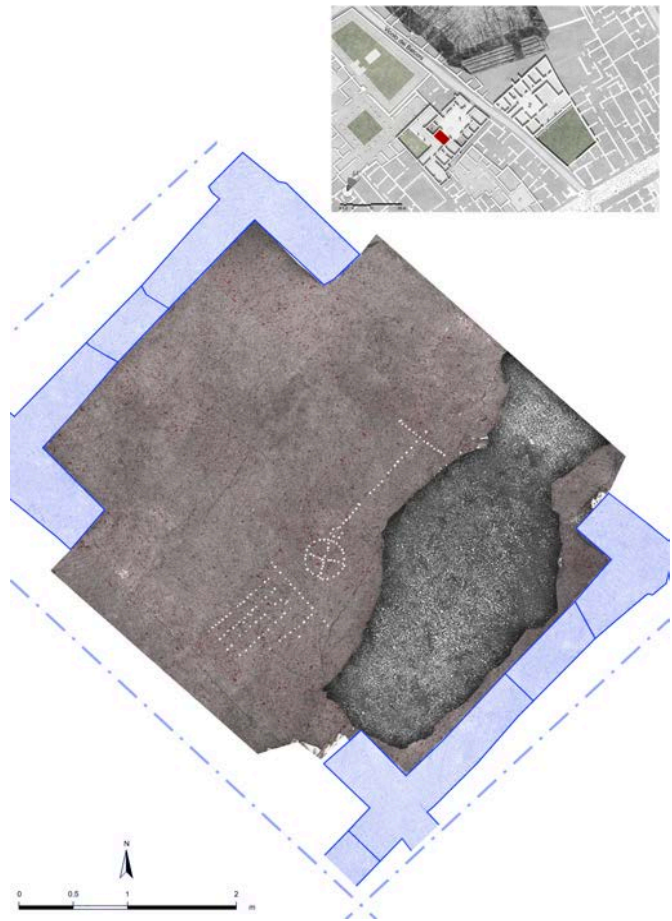

**Figure 4.** The room facing *Atrium* with IMG.2 and IM3.

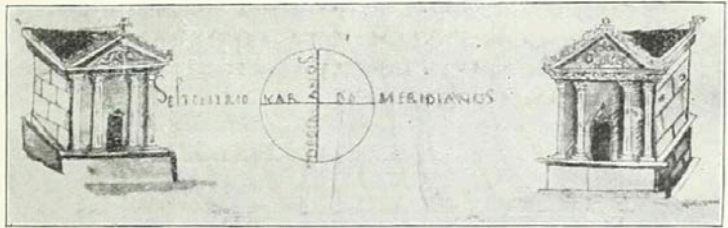
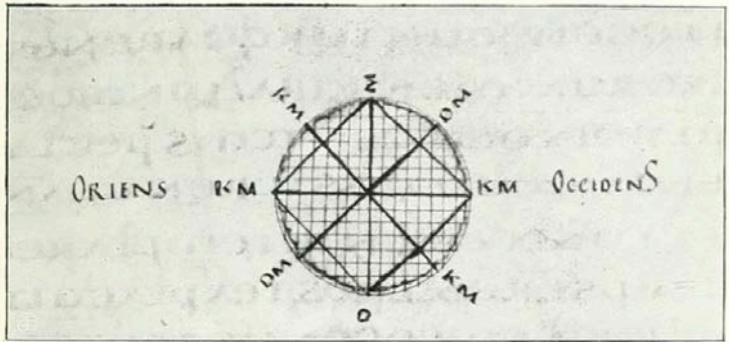
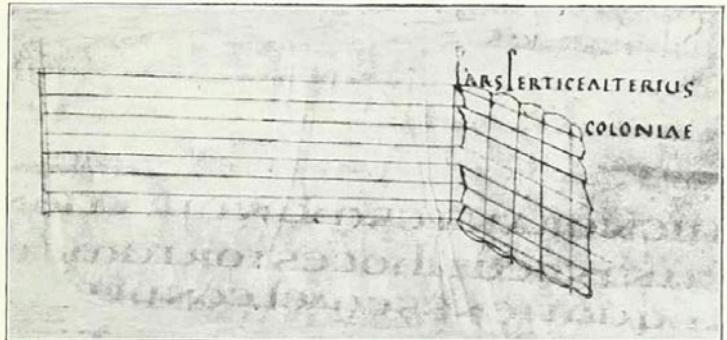

**Figure 5.** Images from the *Corpus* by IGINO.

## 5. ACKNOWLEDGEMENTS

Images 1-4 are by the Authors ©Ferro/Magli/Osanna, Image 5 is in the public domain.   particular thanks to the Architects M. Saldarini and M. Mangini who collaborated to the editing of the images.